\documentclass[useAMS,usenatbib,onecolumn]{mn2e}
\usepackage{epsfig}
\usepackage{amssymb}

\def\aa{{\it Astron. Astrophys.} \,}

\def\apj{{\it ApJ \,}}
\def\apjs{{\it Ap. J. Supp.} \,}

\def\jca{{\it J. Cosmo. Astro. Phys.} \,}

\def\plb{{\it Phys. Lett. B.} \,}
\def\prd{{\it Phys. Rev. D.} \,}
\def\mn{{\it MNRAS} \,}

\title[S-Z power in non-Gaussian and early dark energy models]{Cluster 
abundances and S-Z power spectra: effects of non-Gaussianity and early 
dark energy}
\author[S. Sadeh, Y. Rephaeli, J. Silk]{Sharon Sadeh$^{1}$\thanks{E-mail:
shrs@post.tau.ac.il}, Yoel Rephaeli$^{1,2}$ and Joseph
Silk$^{3}$\\
$^{1}$School of Physics and Astronomy, Tel Aviv University, Tel Aviv, 69978, 
Israel\\
$^{2}$Center for Astrophysics and Space Sciences, University of California,
San Diego, La Jolla, CA 92093-0424\\
$^{3}$Department of Astrophysics, University of Oxford, Keble Road, 
OX1 3RH, UK}

\begin{document}
\pagerange{\pageref{firstpage}--\pageref{lastpage}} \pubyear{2007}

\maketitle

\label{firstpage}

\begin{abstract}
In the standard $\Lambda$CDM cosmological model with a Gaussian primordial 
density fluctuation field, the relatively low value of the mass variance 
parameter ($\sigma_8=0.74^{+0.05}_{-0.06}$, obtained from the WMAP 3-year 
data) results in a reduced likelihood 
that the measured level of CMB anisotropy on the scales of clusters 
is due to the Sunyaev-Zeldovich (S-Z) effect. 
To assess the feasibility of producing higher levels of S-Z power, we 
explore two alternative models which predict higher cluster abundance. 
In the first model the primordial density field has a $\chi^2_1$ 
distribution, whereas in the second an early dark energy component gives 
rise to the desired higher cluster abundance. We carry out the necessary 
detailed calculations of the levels of S-Z power spectra, cluster number 
counts, and angular 2-point correlation function of clusters, and compare 
(in a self-consistent way) their predicted redshift distributions. Our 
results provide a sufficient basis upon which the viability of the three 
models may be tested by future high quality measurements.
\end{abstract}


\section{INTRODUCTION}

Measurements with the BIMA (Dawson et al. 2002), CBI (Readhead et al. 2004), 
and ACBAR (Kuo et al. 2004) experiments indicate a significant power excess 
at high multipoles ($\ell>5000$ in the former, $\ell\sim 3000$ in the two 
latter cases) with respect to theoretically predicted levels of the primary 
CMB anisotropy. It has been suggested that the excess could be due to the S-Z 
effect, but its observed level would require a rather high value of 
$\sigma_8$ ($\gtrsim 1$) in order to have the necessary large number of 
massive clusters, with the largest relative contribution to the S-Z power 
spectrum. Such a high value is usually disfavoured in cluster studies, 
and is at clear variance with results from the latest WMAP 3-year data, 
according to which $\sigma_8=0.74^{+0.05}_{-0.06}$ (Spergel et al. 2006). 
If the reported excess is indeed due to the S-Z effect, modifications to 
the standard $\Lambda$CDM cosmological model may be required in order to 
account for the implied higher number of clusters.

Cluster abundance at high redshift can be boosted by the presence of a 
scale-dependent, non-Gaussian, positively skewed component in the primordial 
density fluctuation field, as was demonstrated by Mathis, Diego, \& Silk 
(2004), and Sadeh, Rephaeli, \& Silk (2006) (hereafter SRS). In such a model, 
primordial overdensities attain larger amplitudes with higher probabilities 
than the corresponding overdensities in a purely Gaussian random field, and 
may therefore give rise to earlier cluster formation and to higher numbers 
of massive clusters. In both works the former WMAP 1-year normalisation was 
employed ($\sigma_8=0.9$), and it was shown that already with this relatively 
high value it is difficult to reconcile the CMB power excess with the 
inferred cluster population, if the underlying fluctuation field obeys 
Gaussian statistics. On the other hand, a $\chi^2_1$ distributed field was 
demonstrated to be capable of producing a sufficiently large cluster 
population, and consequently S-Z power levels that are consistent with 
the above mentioned observational results. 
The $\chi^2_m$ family of models was originally proposed 
by Peebles (1997, 1999a,b), and is based on an isocurvature cold dark matter 
(CDM) scenario in which the primordial density field is proportioinal to the 
square of a random Gaussian process (Peebles 1997). Implications of this 
model on the primary CMB anisotropy and the dynamics and evolution of the 
large scale strcuture have been the subject of several studies (e.g. Koyama, 
Soda \& Taruya 1999, Mathis, Diego \& Silk 2004). Other theoretically 
motivated non-Gaussian models found in the literature include, e.g., the 
log-normal probability distribution function. A rather comprehensive list 
of such models and their statistical properties are explored in detail in 
Coles \& Barrow (1987). A common parametrization of non-Gaussianity is 
provided by the following transformation on an underlying Gaussian field:
\begin{equation}
\psi(x)=\alpha\phi(x)+\epsilon\left[\phi^2(x)-\langle\phi^2\rangle\right]. 
\end{equation}
The parameter $\epsilon$, alternatively written as $f_{NL}$ and referred to 
as the nonlinear coupling factor, characterises the amplitude of the 
quadratic term. In this respect, the $\chi^2_m$ model constitues a special 
case in which $\alpha=0$.  
The degree of non-Gaussianity of a given model is defined in several ways; 
these include the skewness of the density field (for which a one-to-one 
correspondence with the nonlinear coupling term can be derived; Matarrese, 
Verde \& Jimenez 2000), and the parameter $\zeta$ defined as
\begin{equation}
\zeta(M)=\frac{\int_{3\sigma}^{\infty}p(\delta,M)d\delta}
{\int_{3\sigma}^{\infty}p_G(\delta,M)d\delta},
\end{equation}
which describes the 3-$\sigma$ tail excess probability of the non-Gaussian 
model over the Gaussian probability distribution function. For the $\chi^2_1$ model $\zeta=16.3$. While there seems to be 
observational evidence for a lower level of non-Gaussianity ($\zeta\lesssim 
4$, Robinson \& Baker 2000; Avila-Reese et al. 2003, and $\zeta\lesssim 1.6$, Robinson, Gawiser, \& Silk 2000; Komatsu et al. 2003), we believe 
that these results do not necessarily rule out models 
with higher levels of non-Gaussianity. This is due to the fact that 
these results were inferred either from primary CMB anisotropy observations, 
which explore scales much larger than those associated with clusters of 
galaxies, and therefore, lower overdensities, which may be indistinguishable 
from a non-Gaussian distribution in case the density field is scale-variant, 
or from observations of low-redshift clusters, whereas - as will be shown below - the non-Gaussian tail would be mainly manifested in an enhanced high-redshift cluster population.

Early dark energy (hereafter EDE) models provide an alternative for 
generating higher cluster populations at higher redshifts. In these models 
the dark energy density is appreciable already at early epochs of cosmic 
evolution, and attains the observationally inferred value at present. 
Consequently, the quintessence equation of state coefficient changes with 
time. Such models have been the subject of study by several authors who 
investigated their potential influence on the CMB (Doran et al. 2001a, 
Caldwell et al. 2003), and the linear growth of structure (Ferreira \& Joyce 
1998, Doran et al. 2001b). More recently, Bartelmann, Doran, \& Wetterich 
(2006; hereafter BDW) have carried out a comprehensive study of two specific 
EDE models, evaluating numerically the quantities relevant to structure 
formation and the halo mass function, such as the linear growth factor 
of density perturbations, critical density for spherical collapse, and the 
overdensity at virialisation. Specifically, they find that with respect to 
their corresponding values in the standard $\Lambda$CDM model, a slower 
evolution of the linear growth factor and reduced values of the 
critical density for spherical collapse are predicted. As noted by BDW, the 
slower evolution of the linear growth factor in EDE models is a consequence 
of the higher expansion rate of the universe at early times due to an early 
acceleration phase caused by the non-vanishing dark energy component. 
Therefore, with a given value of the present mass variance normalisation 
the corresponding quantity at early times should be larger than what is 
implied in the $\Lambda$CDM model. Since the critical overdensity for 
collapse at a given redshift is the linearly extrapolated value from the 
early universe, the slower evolution of the growth factor in EDE models is 
manifested by a lower critical density with respect to the $\Lambda$CDM model 
at all redshifts relevant to structure formation. Hence, the cluster 
population, which increases with decreasing $\delta_c$ and increasing 
$\sigma_{M}$, grows considerably.

In this paper we focus on the impact of these two alternative scenarios on 
S-Z observables, namely the angular power spectrum of CMB temperature 
variations due to the S-Z effect, and the angular 2-point correlation 
function (hereafter A2PCF) of clusters. As is demonstrated below, 
the higher abundance of massive clusters at high redshifts in the 
non-Gaussian and EDE models boosts S-Z power levels (as a result of higher 
values of the Comptonization parameter), which in turn gives rise to larger 
temperature variations in the CMB. Conversely, the A2PCF is demonstrated to 
predict lower correlation levels in the EDE, and in particular, the 
non-Gaussian model, owing to the detailed properties of the linear bias 
parameter, which, with respect to the $\Lambda$CDM model, has a different 
mass and redshift dependence in the EDE model, and a distinct functional 
form in the non-Gaussian model. 

In \S 2 we briefly describe the Press \& Schechter mass function variant 
adapted to the $\chi^2_1$-distributed probability distribution function 
(PDF) of the primordial density fluctuation field, and outline the properties of the EDE model adopted in our calculations. Results of the calculations are 
described in \S 3, where S-Z power levels calculated for the standard 
$\Lambda$CDM, EDE, and non-Gaussian models are compared with 
the BIMA, CBI, and ACBAR observational results, and an account of the A2PCF 
for these 3 models is given. \S 4 includes a discussion and our conclusions. 
In the appendix we provide a brief exposition of the differential equations 
governing the redshift evolution of the comoving radial distance, linear 
growth factor of density perturbations, critical density for spherical 
collapse, and overdensity at virialisation corresponding to the EDE model, 
and present some numerical results.

\section{METHOD}

The standard $\Lambda$CDM, EDE, and non-Gaussian models will be referred to 
hereafter as models I, II, and III, respectively. A full account of models I 
and III is provided in SRS; here we only include new relevant aspects of 
these two models. We adopt a Press \& Schechter (1974) mass function of the 
form 
\begin{equation}
n(M,z)=-\mu F(\mu)\frac{\rho_b}{M\sigma_M^2(z)}\frac{d\sigma}{dM}\,dM,
\end{equation}
where $\mu\equiv\delta_c(z)/\sigma_M(z)$ is the critical overdensity for 
collapse in terms of the mass variance $\sigma_M$ at redshift $z$, and 
$\rho_b$ is the background density at $z=0$. For models I and II,
\begin{equation}
F(\mu)=\sqrt{\frac{2}{\pi}}e^{-(\mu^2/2)}, 
\end{equation}
whereas the corresponding expression for model III, for which we take a 
primordial density fluctuation field obeying $\chi^2_1$ statistics, is
\begin{equation}
F(\mu)=\frac{e^{-(1+\sqrt{2}\mu)/2}}{\mbox{erfc}[1/\sqrt{2}]\sqrt{\pi(1+\sqrt{2}\mu)}}.
\end{equation}
The factor $\mbox{erfc}[1/\sqrt{2}]$ (with $\mbox{erfc}$ representing 
the complementary error function) in the denominator is introduced in order 
to arrange that all the mass in the universe be included in haloes, 
as is similarly accomplished by introducing the well known factor of 2 in 
the original Press \& Schechter mass function. (Recall that the full 
theoretical justification for this factor is somewhat uncertain; here it 
is merely introduced to ensure consistency with the mass function of model I). The predicted statistical properties of S-Z observables (i.e. weighted by the 
cluster population) will be affected by the model of the mass function; 
clearly, model III differs from models I and II in the functional form of 
the mass function. Model II differs from both models I and III in the 
redshift dependence of the critical density for collapse and the linear 
growth factor of density perturbations, as will be demonstrated below. 

For the calculation of the mass variance $\sigma_R$ we used a top-hat window
function and CDM transfer functions taken from Bardeen et al. (1986): 
adiabatic transfer function for models I and II, and isocurvature transfer 
function for model III. Note that the presence of EDE is likely to affect 
the shape of the CDM transfer function; this effect is minor, however,
with a largest difference of $\lesssim 7\%$ at the lowest wave numbers 
($k\sim 10^{-4} Mpc^{-1}h$), becoming negligible at higher wave numbers 
(which correspond to cluster scales), as was verified using the latest 
version of the CMBFAST code (Zaldarriaga \& Seljak 2000), which admits 
a time-dependent equation of state for dark energy. Consequently, it is 
reasonable to use the fit provided by Bardeen et al. even in the case of 
model II. 

The calculations were carried out using cosmological parameters deduced from 
the WMAP 3-year data: 
$\Omega_{\Lambda}=0.76$ (which, for the case $w\ne -1$ is usually written 
as $\Omega_Q$), $\Omega_m=0.24$, $h=0.73$, and
$\sigma_8=0.74^{+0.05}_{-0.06}$. In addition to these parameters we take the 
spectral index to be $n=1$ in models I and II, and $n=-1.8$ in model III. 
Model II requires additional parametrization of the EDE. We 
use a variant of the EDE models studied in BDW, where we take the density of 
early quintessence with respect to the critical density to be 
$\Omega_e=.0008$, and the equation of state coefficient at $z=0$ to be 
$w_0=-0.99$. The effective coefficient as a function of redshift is given 
in the expression (Wetterich 2004)
\begin{equation}
\overline{w}(z)=\frac{w_0}{1+u\log{(1+z)}},
\end{equation}
where
$u\equiv\frac{-3w_0}{\log{\left(\frac{1-\Omega_e}{\Omega_e}\right)}
+\log{\left(\frac{1-\Omega_m}{\Omega_m}\right)}}$.

The differential equations governing the evolution of the linear growth 
factor, critical density for spherical collapse, and overdensity at 
virialisation pertaining to this model, as well as their numerical solutions, 
are presented in the Appendix. It is important to note that while our choice 
of parameters in model I is completely consistent with WMAP results, this is 
not necessarily the case in models II and III. Nevertheless, we believe that 
we are able to account for this apparent inconsistency by choosing to 
normalise the resulting mass functions so as to yield the same (cumulative) 
cluster density at $z=0$, as predicted by model I. This is justified 
given the fact that the observables quantified in this study are all related 
to clusters of galaxies. In fact, it turns out that the same normalisation of $\sigma_8=0.74$ (in accordance with WMAP results) induces an identical 
cumulative cluster density population at $z=0$, as can be clearly seen by 
inspection Fig.~\ref{fig:mf}. We include for reference a fourth model, which 
is basically identical to model I, but with a different normalisation of 
$\sigma_8=0.8$. This higher value has been advocated by Evrard et al. (2007), 
who claim that with a low normalisation, the specific energy in the IC gas 
and galaxies must increase by $50\%$, which would imply high biases with 
respect to the cluster dark matter, at variance with what is observed in 
hydrodynamical simulations. According to Evrard et al., a higher 
normalisation can solve this conflict, and also provides a better match to 
recent S-Z observations. These results are actually in line with several cluster studies which seem to indicate higher normalisations than inferred from CMB observations. 

The S-Z angular power spectrum was calculated using a formalism similar 
to that adopted by (e.g.) Cooray (2000), Komatsu \& Seljak (2002), and 
Refregier \& Teyssier (2002). 
Note that these authors have used the NFW formalism to characterise the 
cluster dark matter profile and the implied IC gas profile. In our 
calculations we chose to model the IC gas with an isothermal $\beta$-King 
profile with temperatures derived within the framework of hydrostatic 
equilibrium. The reason for our selection stems from the fact that the 
redshift and mass modeling of the concentration parameter appearing in the 
NFW profile have been carried out on the basis of results from N-body 
simulations, which usually incorporate primordial Gaussian fields. In 
such simulations clusters show up only at relatively low redshifts, such 
that it is not clear whether the deduced fitting of the concentration 
parameter can really address also higher-redshift clusters, a natural 
outcome of positively skewed non-Gaussian models. Furthermore, fitting 
formulae of $c(M,z)$ found in the literature fail to yield meaningful 
values at redshifts at which clusters may already form in non-Gaussian 
models. Calculation of the A2PCF of (S-Z) clusters follows the work of SRS and that of Mei \& Bartlett (2003); full details can be found in these papers. 

\section{RESULTS}

\begin{figure}
\centering
\epsfig{file=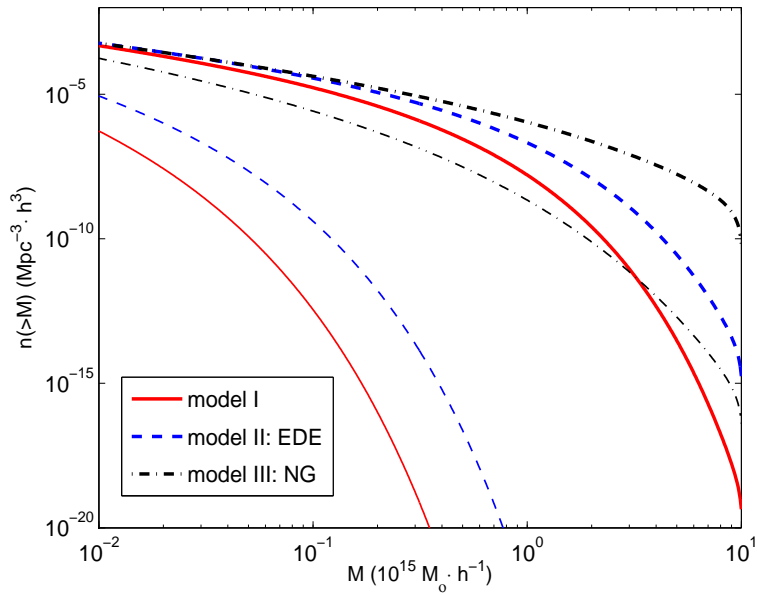, height=8.5cm, width=11cm, clip=}
\caption{Cumulative mass functions for models I (continuous), II (dashed), 
and III (dot-dashed), as a function of cluster mass. The two sets of upper and lower curves correspond to redshifts $z=0.01$ and $z=3$, respectively.}
\label{fig:mf}
\end{figure}

Before we present results of the S-Z power spectrum and A2PCF, it is useful 
to investigate the behaviour of the mass function in the 3 models at various 
redshifts. Fig.~\ref{fig:mf} depicts the cumulative mass function of 
clusters with masses lying in the range $10^{13}M_{\odot}h^{-1}\le M\le 
10^{16}M_{\odot}h^{-1}$ at redshifts $z=0.01$ and $z=3$. As can be 
clearly seen, the three models are correctly normalised so as to yield the 
same cumulative cluster density at low redshifts. As expected, models II and 
III consistently predict higher abundances of high-mass clusters than the 
corresponding high-mass population in the standard $\Lambda$CDM model, with 
model III dominating over model II. This occurs at both low and high 
redshifts, and becomes more pronounced with increasing redshift. 
It appears then that the presence of an excess of high overdensity 
fluctuations associated with the $\chi^2_1$ model has a stronger impact on 
the population of massive clusters than the slower evolution of the linear 
growth factor and lower values of the critical density for collapse 
that characterise model II.

\begin{figure}
\centering
\epsfig{file=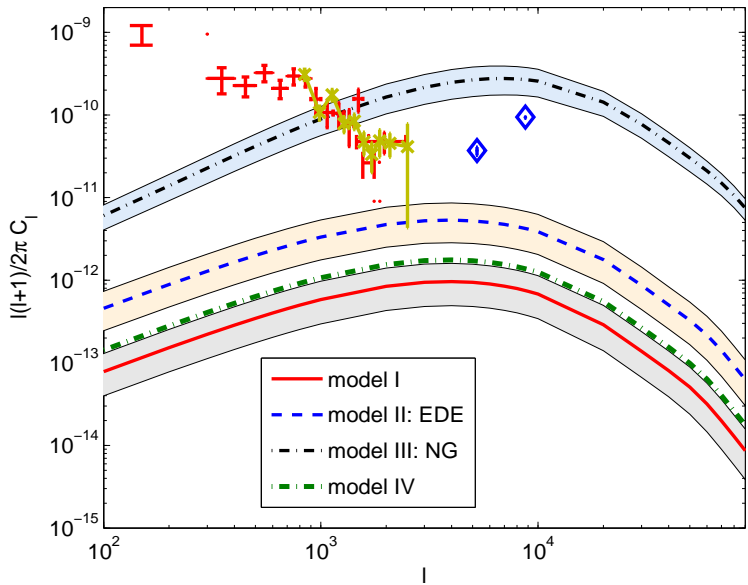, height=8.5cm, width=11cm, clip=}
\caption{S-Z angular power spectrum at $\nu=31\,GHz$ in 
models I (continuous), II (dashed), III (dash-dotted), and IV (thick 
dash-dotted). The shaded areas correspond to the WMAP reported 1-$\sigma$ 
errors in $\sigma_8$. Also shown is the power excess reported by the BIMA 
(diamonds), CBI (crosses) and ACBAR (x-symbols) experiments. Note that the 
BIMA indicated power at $\ell=8748$ is an upper limit.}
\label{fig:cl}
\end{figure}

The behaviour of the respective mass functions is directly reflected in 
the S-Z power spectra illustrated in Fig.~\ref{fig:cl}. Here the enhanced 
massive cluster population, particularly at relatively high redshifts, 
is manifested in both increased power levels
($\sim 1\cdot 10^{-12}, 5\cdot 10^{-12}, 3\cdot 10^{-10}$ in models I, II, 
and III, respectively), and a shift of the peak power towards higher 
multipoles from $\ell\sim 4000$ for models I and II, to $\sim 7000$ for 
model III, reflecting the higher abundance of distant, low angular size 
clusters, particularly in model III. It can also be seen that model 
IV predicts power levels which are a factor $1.8-2$ higher than those 
obtained for model I. The numerical results for the four models were scaled 
to a frequency of 31 GHz (at which the non-relativistic spectral distortion 
due to the thermal S-Z effect amounts to a multiplicative spectral factor of 
$\sim 3.8$ in the expression for the power spectrum) so as to correspond to 
BIMA, CBI, and ACBAR observational results (that are also) shown in the 
figure. Of the four models, model III provides the best match to the observed 
CMB power excess, particularly so at $\ell\sim 1000-1500$ (at higher 
multipoles the calculated power levels are actually higher than the CBI and 
ACBAR results), whereas the predicted power in model I is far below and seems 
clearly inconsistent with the observed level. Although the predicted power 
levels in model II are higher than in model I, these are still only 
marginally consistent with the observational results. 
    
\begin{figure}
\centering
\epsfig{file=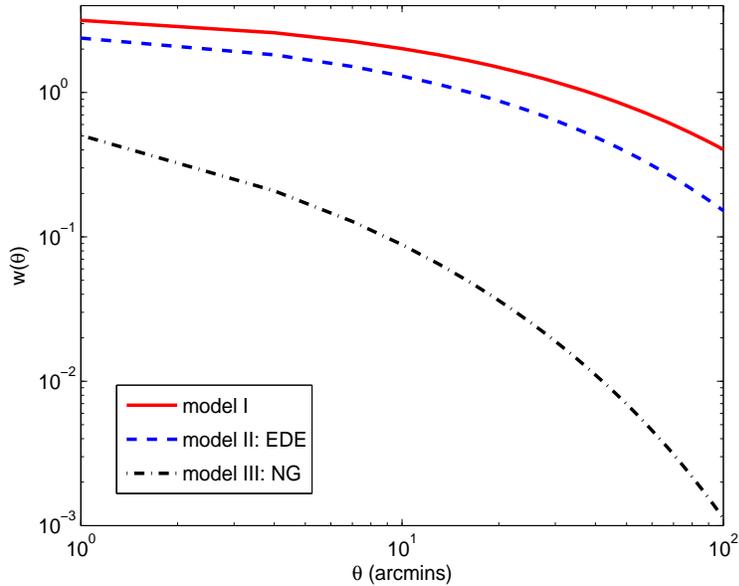, height=8.5cm, width=11cm, clip=}
\caption{The A2PCF of clusters calculated from their S-Z fluxes in 
models I (continuous), II (dashed), and III (dash-dotted), as a function of 
separation angle, $\theta$. The calculations were carried out for the 
$353\,GHz$ channel of the Planck/HFI experiment, with a beamsize of 
$5.1'$ and a limiting flux of $30\,mJy$.} 
\label{fig:corr}
\end{figure}

While both the EDE and non-Gaussian models generate higher S-Z power levels 
with respect to those predicted by the standard $\Lambda$CDM model, the 
levels of the A2PCF of S-Z clusters are lower in model II, and particularly 
so in model III, with respect to the corresponding levels predicted by 
model I, as illustrated in Fig.~\ref{fig:corr}. The A2PCF of models I, II, 
and III peak at $w\sim 2, 1, 0.5$ at an angular separation of $1'$, falling 
off towards levels of $w\sim 0.4, 0.2, 0.001$ at angular separation $100'$. 
Also noticeable is the fact that the curve associated with model III has a 
greater slope than the corresponding slopes of models I and II. In order to 
understand these properties, it is useful to consider the redshift and mass 
dependence of the linear bias factor, which in the Gaussian case assumes the 
form (Mo \& White 1996):
\begin{equation}
b(M,z)=1+\frac{\mu^2-1}{\delta_c(z)},
\label{eq:bias1}
\end{equation}
where $\mu\equiv\delta_c(z)/[\sigma_M(0)D_{+}(z)]$, $D_{+}(z)$ 
denotes the linear growth factor of density perturbations, and $\sigma_M(0)$ 
is the mass variance at $z=0$. Koyama, Soda, \& Taruya (1999) generalised 
this expression to also address non-Gaussian models. Implementing their 
expression to the $\chi^2_1$ model, we have
\begin{equation}
b(M,z)=1+\frac{\mu^2-1}{\delta_c(z)[1+\sqrt{2}\mu]}.
\label{eq:bias2}
\end{equation}
The lower bias levels of massive haloes in non-Gaussian models are a 
consequence of the linear bias theory developed by Mo \& White (1996), and 
were explored in detail by Amara \& Refregier (2004) within the framework of 
the log-normal PDF. In fact, these authors showed that for a log-normal PDF 
with an excess probability in the $3\sigma$ tail of $\zeta\sim 10$, the 
corresponding bias factor is larger than the one predicted by the 
Gaussian model at halo masses $\gtrsim 5\cdot 10^{13}\,M_{\odot}\,h^{-1}$. 
Since the non-Gaussian model considered here has 
$\zeta\sim 16$, the mass scale at which the bias factor begins to be 
dominated by the corresponding Gaussian bias reduces to 
$\sim 10^{13}\,M_{\odot}\,h^{-1}$, which is the lower limit of our mass 
integral. It is therefore safe to assume that the 
relevant bias factor is lower than that of the Gaussian model within the entire mass range.

\begin{figure}
\centering
\epsfig{file=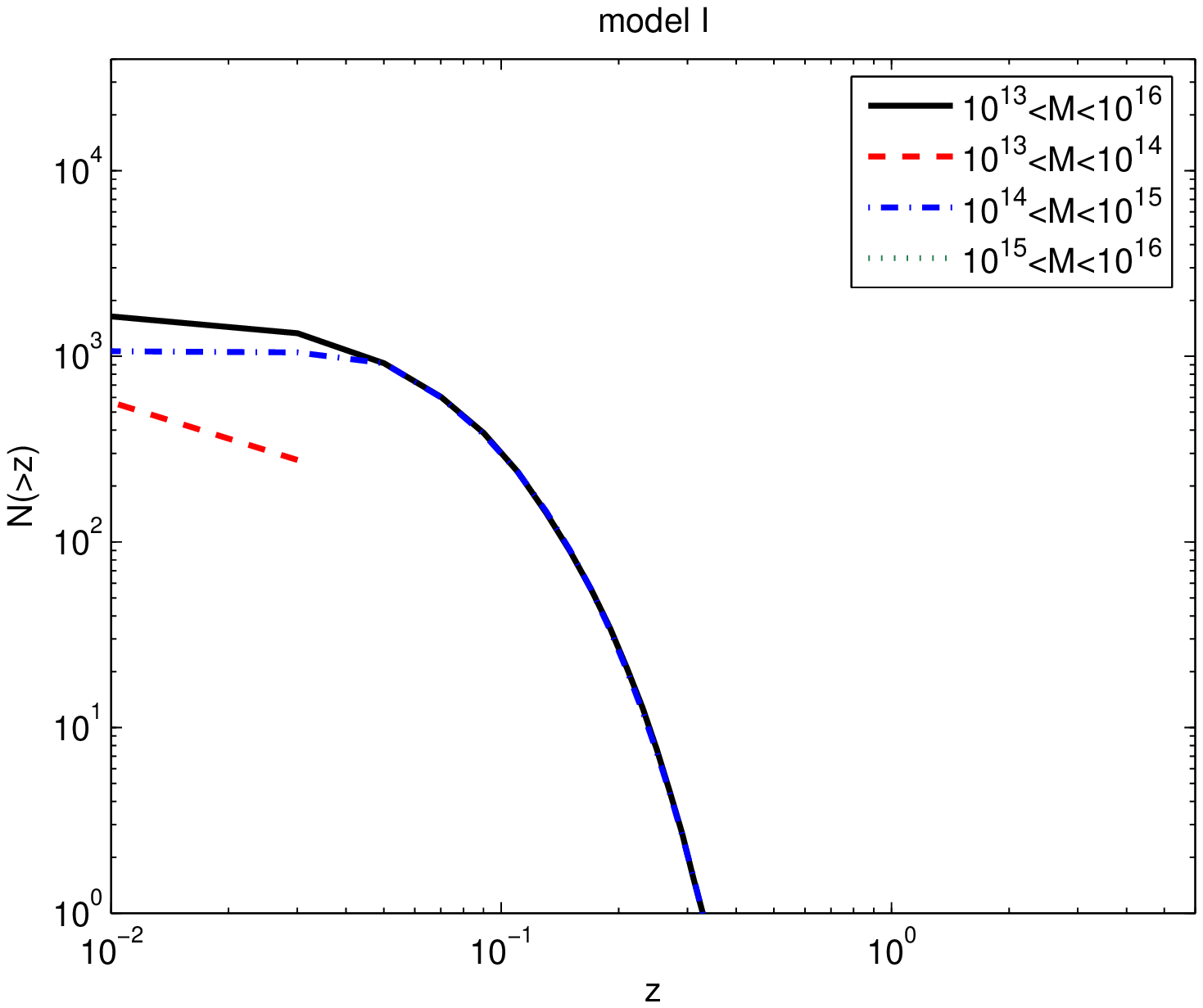, height=8cm, width=8.8cm, clip=}
\epsfig{file=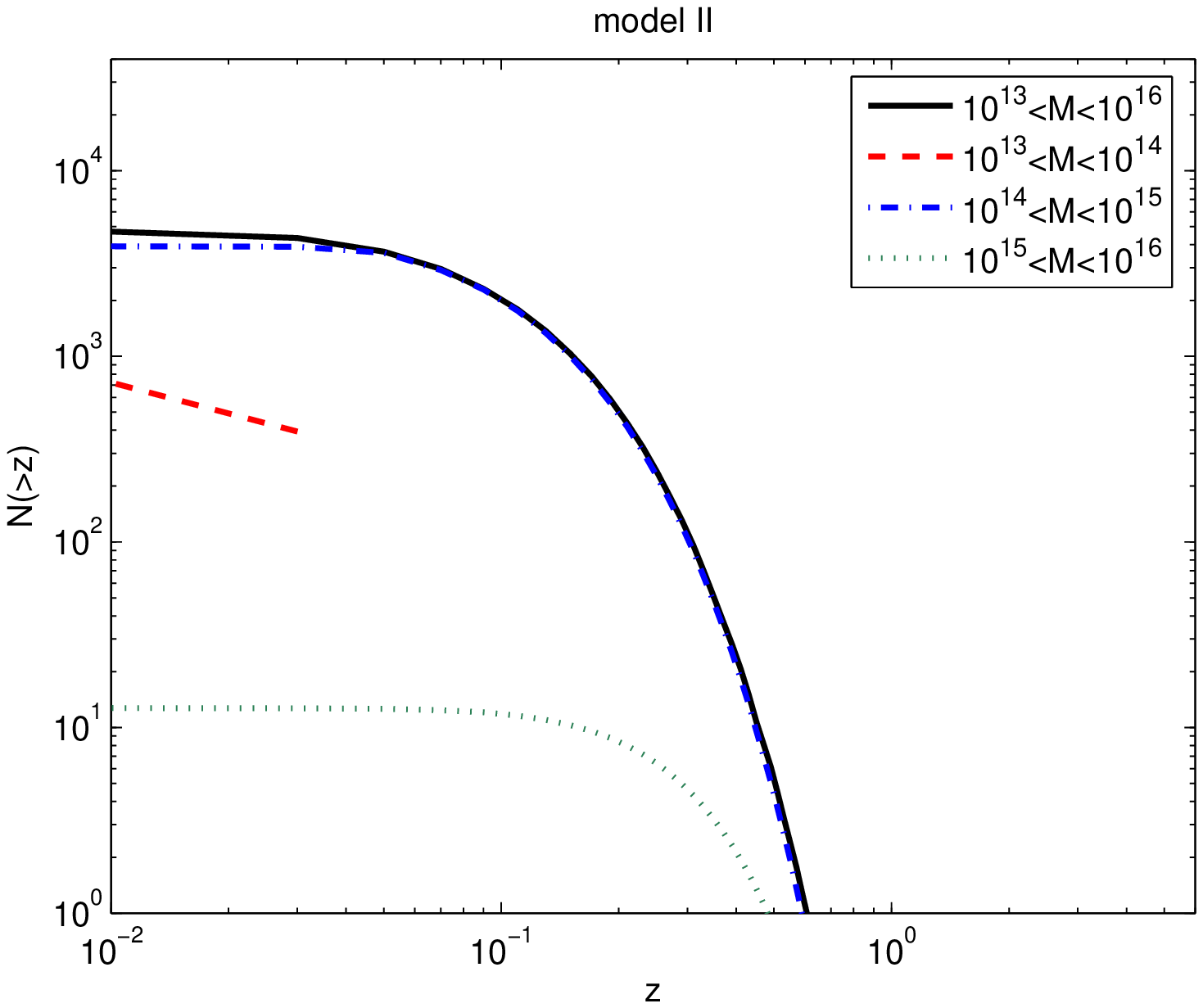, height=8cm, width=8.8cm, clip=}
\epsfig{file=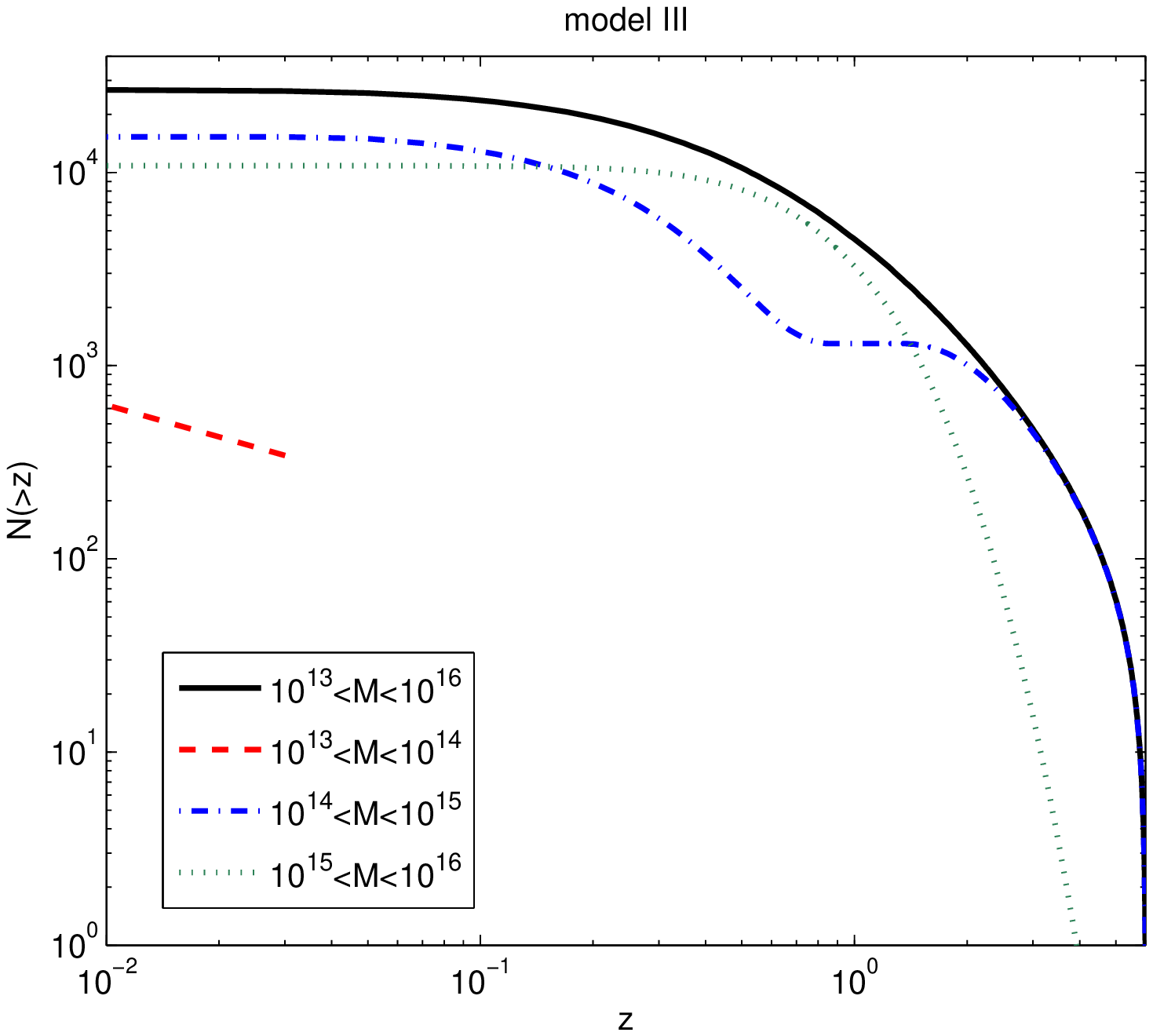, height=8cm, width=8.8cm, clip=}
\caption{Cumulative number counts of S-Z clusters with fluxes exceeding 
$30\,mJy$ as a function of redshift. The calculations were carried out 
at a frequency of $353\,GHz$ and beam size $5.1'$. Upper left- and 
right-hand panels and the bottom panel correspond to models I, II, and III, 
respectively. Continuous curves depict the contribution to the cumulative 
number counts in the entire mass range of 
$10^{13}M_{\odot}h^{-1}\le M\le 10^{16}M_{\odot}h^{-1}$; dashed, 
dash-dotted and dotted curves represent contributions from partial mass 
ranges, as indicated in the legend.}
\label{fig:cnc}
\end{figure}

The properties of the A2PCF can be now easily explained by inspection of the 
functional dependence of the bias factor on mass and redshift. The bias 
factor increases with increasing mass, since the mass variance decreases 
monotonically with increasing mass. Additionally, the bias factor increases 
with increasing redshift since the critical density for collapse increases 
with increasing redshift, whereas the linear growth factor decreases with
increasing redshift. These properties are common to all of the three models. 
The differences between the predictions of the 3 models arise either from 
the different evolution of $\delta_c(z)$ and $D(z)$ in model II, or the 
distinct functional form of the bias factor in model III. In model II, the 
magnitude of the bias factor is lower than in model I, in light of the lower 
and higher values of the critical density for collapse and linear growth 
factor, respectively (Fig.~\ref{fig:evo}). This is reflected in the lower 
levels of the A2PCF in model II, with respect to model I. More striking are 
the differences between the predicitons of models I and III. As can be easily 
realized from equation~(\ref{eq:bias2}), the bias is reduced owing to the 
addition of a larger than unity factor in the denominator of the second term, 
which is reflected in the substantially lower levels of the A2PCF. 
Note however, that the differences between the correlation levels of models 
I and III increase with increasing angular separation. At low angular 
separations both nearby and distant clusters contribute to the A2PCF by 
virtue of their relatively low mutual distances. Since the occurence of 
high-redshift, massive clusters, capable of generating the required flux 
limit, is substantially larger in model III, these contribute significanly 
to the correlation levels of model III at low angular separations, whereas 
the corresponding contribution in model I originates in clusters residing at 
lower redshifts. At higher angular separations, where correlations can only 
arise locally, a relative deficit of low redshift clusters appear in model 
III, as can be seen in Fig.~\ref{fig:cnc}, where we plot cumulative number 
counts of S-Z clusters (evaluated at a frequency of $353\,GHz$, with beam 
size of $5.1'$, and flux detection limit $30\,mJy$). Indeed, if we compare 
the upper left-hand panel with the lower panel of Fig.~\ref{fig:cnc}, we may 
see that while (cumulative) cluster counts continue to increase in the 
redshift range $0.2\gtrsim z\gtrsim 0.01$ in model I, they level off at 
$z\lesssim 0.2$ in model III. This property of model III gives rise to the 
faster descent of the A2PCF, with respect to model I, at high angular 
separations.   

The results presented so far focused on the differences in S-Z power and angular correlation levels in the three models due to the respective mass functions and bias factors. Internal cluster properties relevant to the S-Z effect, such as virial size and the temperature and density of the intracluster gas, may be affected as well, as can be seen in  
the bottom panel of Fig.~\ref{fig:evo}, where $\Delta_v(z)$, the overdensity (scaled to the critical density) at virialisation is plotted as function of redshift. However, the variation of $\Delta_v(z)$ among the models amounts to less than $4\%$, implying little change in the above cluster internal properties, particularly so in comparison with the major differences apparent in the mass function and bias parameter.   

\section{DISCUSSION}

The reported low value of $\sigma_8$ inferred from the WMAP 3-year data 
implies a significantly reduced population of high-mass clusters of galaxies,
as predicted by a Press \& Schechter mass function based on a Gaussian 
density fluctuation field. With this low value it becomes even more 
difficult to attribute the CMB power excess measured with the CBI and ACBAR experiments at high multipoles to the (thermal) S-Z effect. The reason for this difficulty lies in the fact that the largest contribution to S-Z power 
comes from high-mass clusters, whose number density decreases steeply with 
decreasing $\sigma_8$. If indeed this excess is attributed to the S-Z effect, an enhanced massive cluster population at moderate and high redshifts is required. In this work we have studied the implications of two non-standard cosmological models on S-Z observables, specifically, the angular power spectrum, and A2PCF. The model based on a primordial density fluctuation field obeying $\chi^2_1$ statistics generates a more abundant population of massive clusters by virtue of higher probabilities for overdense regions at high redshifts. These collapse earlier and form cluster haloes, whereas the early quintessence model is characterized by earlier collapse owing to the higher linear growth factor and lower value of the critical density for collapse at high redshifts. Both models give rise to higher S-Z power levels, more so in the non-Gaussian model. 

We note that the two non-standard models investigated here are by no means 
unique; for example, lower S-Z power is expected in non-Gaussian models 
based on $\chi^2_m$ statistics with $m>1$ due to less pronounced skewness (or non-Gaussianity) of the density field. Likewise, early quintessence models with higher early dark energy densities are likely to further slow down the evolution of the linear growth factor (reducing the value of $\delta_c$) and thus give rise to higher power levels. Conversely,  
the A2PCF of S-Z clusters in the EDE model, and to a significantly higher degree in the non-Guassian model, is manifested in reduced correlation levels with respect to those of the standard $\Lambda$CDM model, in particular at high angular separations.  

EDE models with still larger quintessence densities at early times would 
enhance power levels and further blur the distinction between such models 
and non-Gaussian models. On the other hand, the A2PCF provides a more adequate test of the viability of these models. In this regard we mention the recent work of Magliocchetti et al. (2006), who report results from analysis of Spitzer Space Telescope observations that presumably provide evidence for strong clustering of a galaxy population at $z\sim 2$. According to Magliocchetti et al., these observations suggest that the 
detected objects are very massive proto-spheroidal galaxies, with number 
densities that are considerably higher than predicted by theory. If this is indeed the case, it adds to a list of other observational indications of enhanced massive objects lying at relatively high redshifts, among which are the detection of structures with high velocity dispersions at redshifts $\sim 4$ (Miley et al. 2004) and $\sim 2$ (Kurk et al. 2004), and the detection of protoclusters with masses lying in the range $2-9\cdot 10^{14}\, M_{\odot}$ associated with radio galaxies lying at $z>2$ (Venemans et al. 2006). All this observational evidence points to the presence of massive objects at high redshifts, which - as shown here - may be accounted for by either a non-Gaussian, or early quintessence models. If additional evidence (such as reported by Magliocchetti et al.) is found for strong clustering at $z\sim 2$, it will add further support for non-Gaussian models.
It is important to stress that these observational results pointing to strong clustering of galaxies are not necessarily in conflict with the  
lower correlations found for the non-Gaussian model presented in the previous section. This is due to the fact that our calculations include the contribution to the A2PCF from clusters of galaxies alone, for which, as has been explained above, the (non-Gaussian) bias is lower than the corresponding Gaussian bias. At lower masses, such as galaxies, the bias is actually stronger in non-Gaussian models, and would yield higher levels of the A2PCF. Finally, it is noteworthy that the linear bias factor fails to describe the actual clustering on galaxy scales, since these are already considerably non-linear. Models admitting non-linear corrections exist (e.g. Peacock \& Dodds 1996), but are not likely to alter the general behaviour of the bias parameter in the non-Gaussian and EDE models. 

The non-standard cosmological models considered in this paper give rise to enchanced S-Z power by virtue of their impact on the cluster population. A legitimate question in this regard is whether internal cluster properties alone may be able to account for the apparent variance between the low $\sigma_8$ deduced from the WMAP 3-year data and the high-multipole CMB power excess observed by the ACBAR, CBI, and possibly BIMA experiments. We believe, however, that the answer to this question is negative. IC gas properties and their impact on the S-Z power spectrum have been extensively investigated in the literature; specifically, consequences of the gas mass fraction evolution (Majumdar 2001), isothermal, as well as polytropic temperature profiles (e.g. Komatsu \& Seljak 2002), self-similar and entropy-driven models for the cluster core (Komatsu \& Kitayama 1999), and the temperature-mass relation (SRS), were explored, and however significant differences were found, they are still incapable of settling the conflict. 
Recently, Roychowdhury, Ruszkowski, \& Nath (2005) have explored the influence of AGN heating on groups and clusters of galaxies, motivated by the observed entropy excess in the IC gas. A consequence of this heating was shown to be a depletion of gas from the central regions of clusters, resulting in reduced S-Z power levels. Obviously, this cannot provide a remedy to the conflicting low $\sigma_8$ and high-multipole CMB power excess.   

In conclusion we re-iterate the main result of our work: the mounting 
observational evidence for the presence of massive clusters already at 
relatively high redshifts cannot be easily reconciled with the Press \& Schechter mass function (or variants thereof) in the standard $\Lambda$CDM cosmology with underlying Gaussian density fluctuation field. Non-Gaussian fields and early quintessence models may remove the inherent difficulty by allowing a more abundant population of massive clusters at high redshift. 
Moreover, such models give rise to enhanced S-Z power levels, which seem to 
be in better agreement with the CMB power excess at high multipoles observed by the CBI and ACBAR experiments. Future S-Z cluster catalogs will enable the construction of A2PCF from S-Z measurements, thereby providing 
an additional test for the viability of these models.

\section{APPENDIX}
%%%\renewcommand{\theequation}{A-\arabic{equation}}
% redefine the command that creates the equation no.
%%%\setcounter{equation}{0}  % reset counter 
%\section*{APPENDIX}  % use *-form to suppress numbering
%\appendix
%\index{}\section{A}

Extension of the numerical computation of the S-Z angular power spectrum and A2PCF to EDE models necessitates a re-evaluation of several cosmological quantities that are affected by the presence of early dark energy. 
These include the (comoving) radial and angular diameter distances, the 
linear growth factor of density perturbations, the critical density for 
spherical collapse, and the overdensity at virialisation. This Appendix details the necessary modifications to the basic equations governing the 
redshift (or equivalently, the scale factor, $a$) dependence of these functions, and presents their numerical results.

The EDE model studied in this work is characterised by flat geometry, with an early quintessence density specified by the parameter $\Omega_e$, and an effective redshift-dependent coefficient of the equation of state given by 
\begin{equation}
\overline{w}(z)=\frac{w_0}{1-[3w_0\log{(1+z)}]/
(\log{\frac{1-\Omega_e}{\Omega_e}}+\log{\frac{1-\Omega_m}{\Omega_m}})},
\end{equation}
where $w_0=w(z=0)$. By using the redshift scaling of the density of the dark energy component corresponding to this model,
\begin{equation}
\rho_Q=\rho_{Q_0}(1+z)^{3[1+\overline{w}(z)]},
\end{equation}
in the Friedmann equation 
\begin{equation}
\left(\frac{\dot{a}}{a}\right)^2=\frac{8\pi G}{3}(\rho_m+\rho_Q),
\end{equation}
one obtains
\begin{equation}
\frac{\dot{a}}{a}\equiv H=H_0\left[\Omega_m(1+z)^3+(1-\Omega_m)(1+z)
^{3[1+\overline{w}(z)]}\right]^{1/2}.
\label{eq:fried1}
\end{equation}

\vspace{1cm}
\noindent
{\bf{\large{1. Comoving radial and angular diameter distances}}\vspace{1cm}}

\noindent
The comoving radial distance can be obtained from equation~\ref{eq:fried1} 
as
\begin{equation}
r(z)=\frac{c}{H_0}\int_0^{z}\frac{dz'}{\left[\Omega_m(1+z)^3
+\Omega_Q(1+z)^{3[1+\overline{w}(z)]}\right]^{1/2}},
\end{equation}
with $\Omega_Q=1-\Omega_m$, whereas the angular diameter distance is simply 
\begin{equation}
d_A=\frac{r(z)}{1+z}.
\end{equation}

\vspace{.5cm}
\noindent
{\bf{\large{2. Linear growth factor}}\vspace{1cm}}

\noindent
The differential equation governing the evolution of the linear growth 
factor of density perturbations is
\begin{equation}
\frac{d^2 \delta}{da^2}+\left[\frac{2}{a}-\frac{1}{2a}
\frac{\Omega_Q a^{-3[1+\overline{w}(z)]}[1+3\overline{w}(z)]
+\Omega_m a^{-3}}{\Omega_Q a^{-3[1+\overline{w}(z)]}+\Omega_m a^{-3}}\right]\frac{d\delta}{da}-\frac{3}{2}\frac{\Omega_m}{a^5}
\frac{1}{\Omega_Q a^{-3[1+\overline{w}(z)]}+\Omega_m a^{-3}}\delta=0. 
\end{equation}

\vspace{1cm}
\noindent
{\bf{\large{3. Critical density for spherical collapse}}\vspace{1cm}}

\noindent
Using the Friedmann equations for both the spherical overdensity and 
the background manifold, and denoting by $a_{ta}$ and $R_{ta}$ the scale factor and radius of the spherical overdensity at turnaround, respectively, $\rho_{{Qb},{ta}}$, $\rho_{{mb},{ta}}$, and $\rho_{{mc},{ta}}$ the background density of the dark energy component, the background density of the matter component, and the collapsed density of the matter component at turnaround, respectively, we have 
\begin{eqnarray}
&&\left(\frac{\dot{R}}{R}\right)^2=\frac{8\pi G}{3}
(\rho_{mc}+\rho_{Qc})-\frac{k}{R^2} \\
\nonumber \\
&&\left(\frac{\dot{a}}{a}\right)^2=\frac{8\pi G}{3}
(\rho_{mb}+\rho_{Qb}),
\end{eqnarray}
where the curvature $k$ is assumed to vanish in our model, and the indexes 
$c$ and $b$ refer to collapsed and background components, respectively, the following differential equation may be obtained:
\begin{equation}
\left(\frac{dy}{dx}\right)^2=\frac{\zeta(y^{-1}-1)+
\nu(y^{-1}R_{ta}^{3\overline{w}(R_{ta})}/R^{3\overline{w}(y\cdot R_{ta})}-1)}{x^{-1}+\nu x^{-1}a_{ta}^{3\overline{w}(a_{ta})}/(x\cdot a_{ta})^{3\overline{w}(x\cdot a_{ta})}},  
\label{eq:y-x}
\end{equation}
with $x\equiv a/a_{ta}$, $y\equiv R/R_{ta}$, $\nu\equiv \rho_{{Qb},{ta}}/\rho_{{mb},{ta}}$, and $\zeta\equiv \rho_{{mc},{ta}}/\rho_{{mb},{ta}}=\left(R_{ta}/a_{ta}\right)^{-3}$. Since $x=0$ and $y=0$ at $t=0$, whereas $x=1$ and $y=1$ at turnaround, it is possible to recast equation~\ref{eq:y-x} in the form
\begin{eqnarray}
&&\int_0^1\frac{dy}{\left[\zeta(y^{-1}-1)+
\nu(y^{-1}(a_{ta}\zeta^{-1/3})^{3\overline{w}(a_{ta}\zeta^{-1/3})}
/(y\cdot (a_{ta}\zeta^{-1/3}))^{3\overline{w}(y\cdot (a_{ta}\zeta^{-1/3}))}-1)\right]^{1/2}} \nonumber \\
&&=\int_0^1\frac{dx}{\left[x^{-1}+\nu x^{-1}a_{ta}^{3\overline{w}(a_{ta})}/(x\cdot a_{ta})^{3\overline{w}(x\cdot a_{ta})}\right]^{1/2}},
\label{eq:zeta}
\end{eqnarray}
where we have used the relation $R_{ta}=a_{ta}\zeta^{-1/3}$. 
Equation~(\ref{eq:zeta}) can be solved numerically to yield the sought 
value of $\zeta$, and the critical density for spherical collapse can 
then be found as (Zeng \& Gao 2005a):
\begin{equation}
\delta_c=\frac{3}{5}a_{ta}^{-1}\left(\zeta^{1/3}+
\nu\,a_{ta}\zeta^{-2/3}\right)D(a_c),
\end{equation}
where $D(a_c)$ is the linear growth factor at the time of collapse.

\vspace{1cm}
\noindent
{\bf{\large{4. Overdensity at virialisation}}\vspace{1cm}}

The overdensity of the virialised matter component can be found by 
relating the background and collapsed matter densities at turnaround 
with their counterparts at the time of collapse:

\begin{equation}
\Delta_v=\frac{\rho_{mc,c}}{\rho_{mb,c}}=
\frac{\rho_{mc,ta}}{\rho_{mb,ta}}\cdot\frac{\left(R_{ta}/R_c\right)^3}
{\left(a_{ta}/a_c\right)^3}\equiv\zeta\cdot\lambda^{-3}
\left(\frac{a_c}{a_{ta}}\right)^3,
\end{equation}
or if we are rather interested in the virialised overdensity in terms of 
the critical density at the time of collapse
\begin{equation}
\Delta_v^*=\frac{\rho_{mc,c}}{\rho{c,c}}=\Omega_m(a_c)\cdot\Delta_v,
\end{equation}
where we have defined $\lambda\equiv R_c/R_{ta}$. It remains to specify 
this parameter. This can be done by requiring energy conservation at turnaround and virialisation, in conjunction with the virial theorem applied to the virialised phase of the collapsed overdensity. Before proceeding along this line, we remark that there is an ongoing debate regarding the assumption of energy conservation within the collapsing halo in models with dark energy featuring $w\ne -1$ (e.g. Basilakos \& Voglis 2007, Zeng \& Gao 2005b, Maor \& Lahav 2005). In fact, Maor \& Lahav show that in such models the assumption of energy conservation breaks down. 
Here we choose to ignore this issue, and use energy conservation as a zero order approximation. It is also of interest to note that according to Wang (2006), under certain conditions (namely, for $\nu/\zeta<0.01$), the problematics associated with energy conservation do not affect the virialisation process. Although this factor is somewhat larger in our adopted model, we believe that our results are not significantly affected by this ambiguity. 

To proceed, the potential energy stored in the matter component of the 
spherical overdensity is the usual
\begin{equation}
U_G=-\frac{3GM^2}{5R},
\label{eq:ugm}
\end{equation}
whereas the potential energy associated with the dark energy in the spherical overdensity is (Horellou \& Berge 2005)
\begin{equation}
U_Q=(1+3\overline{w})\rho_Q\cdot\frac{4\pi GM}{10}R^2.
\label{eq:ugq}
\end{equation}
At turnaround the spherical overdensity has only potential energy, while 
at virialisation it has both kinetic and potential energies: 
\begin{equation}
(T+U)_{ta}=(T+U)_c.
\end{equation}
Using the virial theorem, $T_c=-1/2U_G+U_Q$, so
\begin{equation}
\frac{1}{2}U_{G,c}+2U_{Q,c}=U_{G,ta}+U_{Q,ta}.
\label{eq:enerb}
\end{equation}
Substitution of equations~(\ref{eq:ugm}) and~(\ref{eq:ugq}) in 
equation~(\ref{eq:enerb}) (and performing some algebraic manipulations) leads to the following equation:
\begin{equation}
\frac{R_{ta}}{R_c}=\frac{4\pi[1+3\overline{w}(a_{ta})]\rho_{Q,ta}R_{ta}^3-6M}
{8\pi[1+3\overline{w}(a_c)]\rho_{Q,c}R_{c}^3-3M}.
\end{equation}
Using the fact that the mass can be represented by either 
$M=4\pi/3\rho_{mc,ta}R_{ta}^3$ or $M=4\pi/3\rho_{mc,c}R_{c}^3$, this can 
be written as 
\begin{equation}
\frac{R_{ta}}{R_c}=\frac{[1+3\overline{w}(a_{ta})]\frac{\Omega_{Q,ta}}
{\Omega_{mb,ta}}\frac{1}{\zeta}-2}{2[1+3\overline{w}(a_c)]
\frac{\Omega_{Q,c}}{\Omega_{mb,c}}\left(\frac{a_{ta}}{a_c}
\right)^3\left(\frac{R_c}{R_{ta}}\right)^3\frac{1}{\zeta}-1}.
\end{equation}
Given the collapse scale factor $a_c$, or, alternatively, the collapse 
redshift $z_c$, all of the density parameters appearing in this equation 
can be found. Substituting $\lambda\equiv{R_c}{R_{ta}}$, the last equation can be written as 
\begin{equation}
\frac{n_1-2}{2n_2\lambda^3-1}=\frac{1}{\lambda},
\end{equation}
or
\begin{equation}
2n_2\lambda^3-(n1-2)\lambda-1=0,
\end{equation}
where $n_1\equiv[1+3\overline{w}(a_{ta})]\zeta^{-1}(\Omega_{Q,ta}/
\Omega_{mb,ta})$, and $n_2\equiv[1+3\overline{w}(a_c)]
\zeta^{-1}(\Omega_{Q,c}a_{ta}^3)/(\Omega_{mb,c}a_c^{3})$ 
can all be evaluated given the collapse redshift. The solution for $\lambda$ 
should obviously lie in the range $0\le\lambda\le 1$.

\begin{figure}
\centering
\epsfig{file=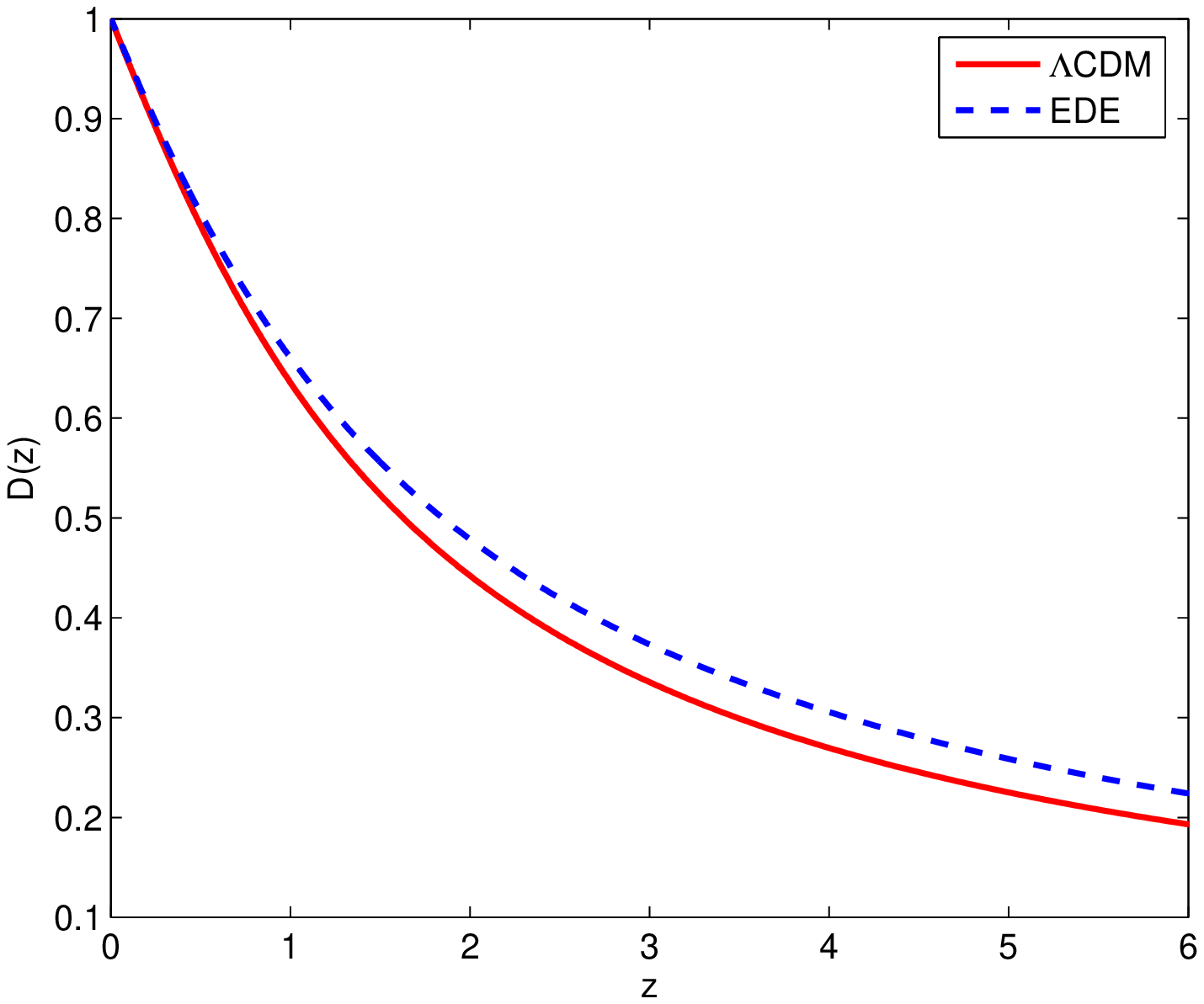, height=7.cm, width=8.cm, clip=}
\epsfig{file=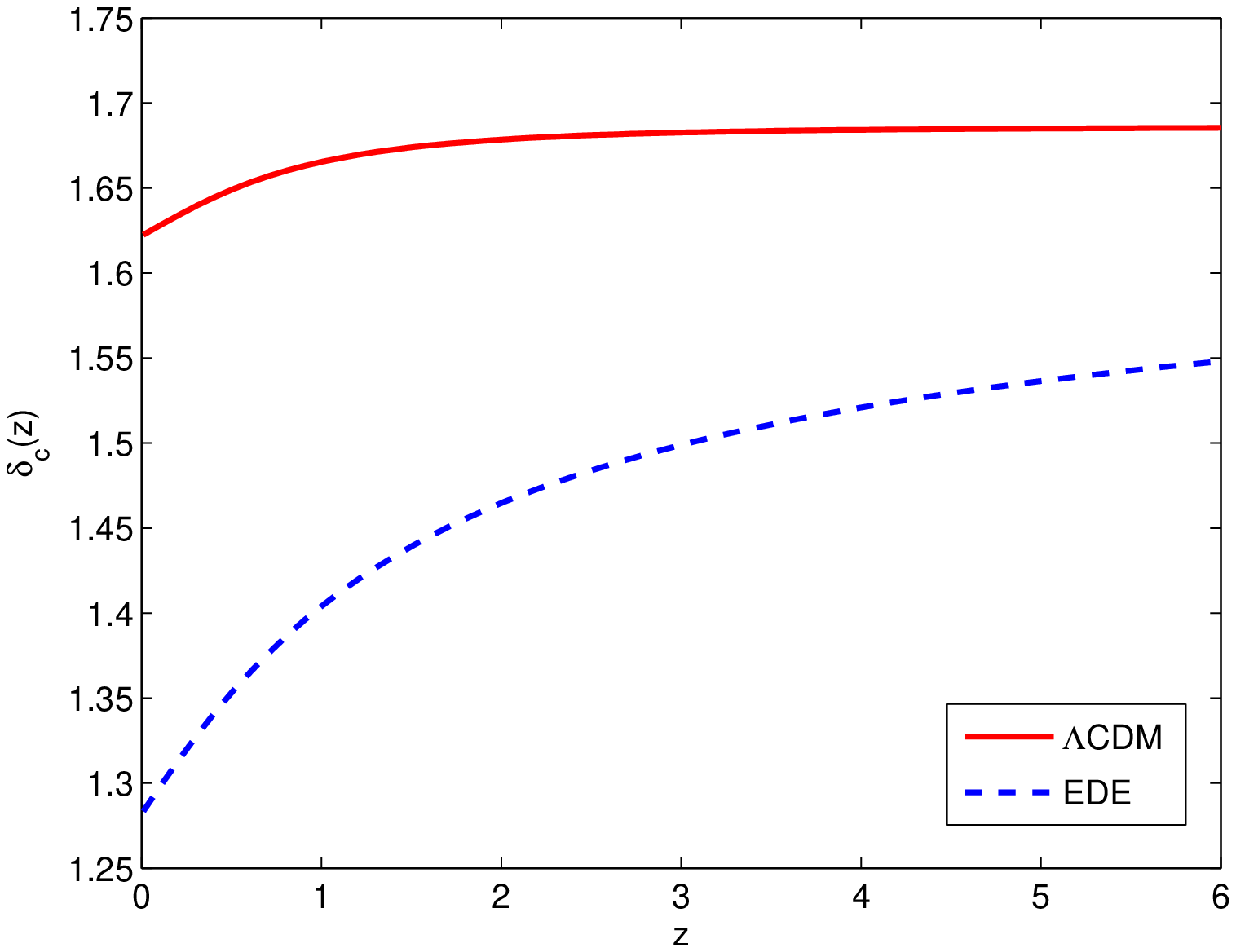, height=7.cm, width=8.cm, clip=}
\epsfig{file=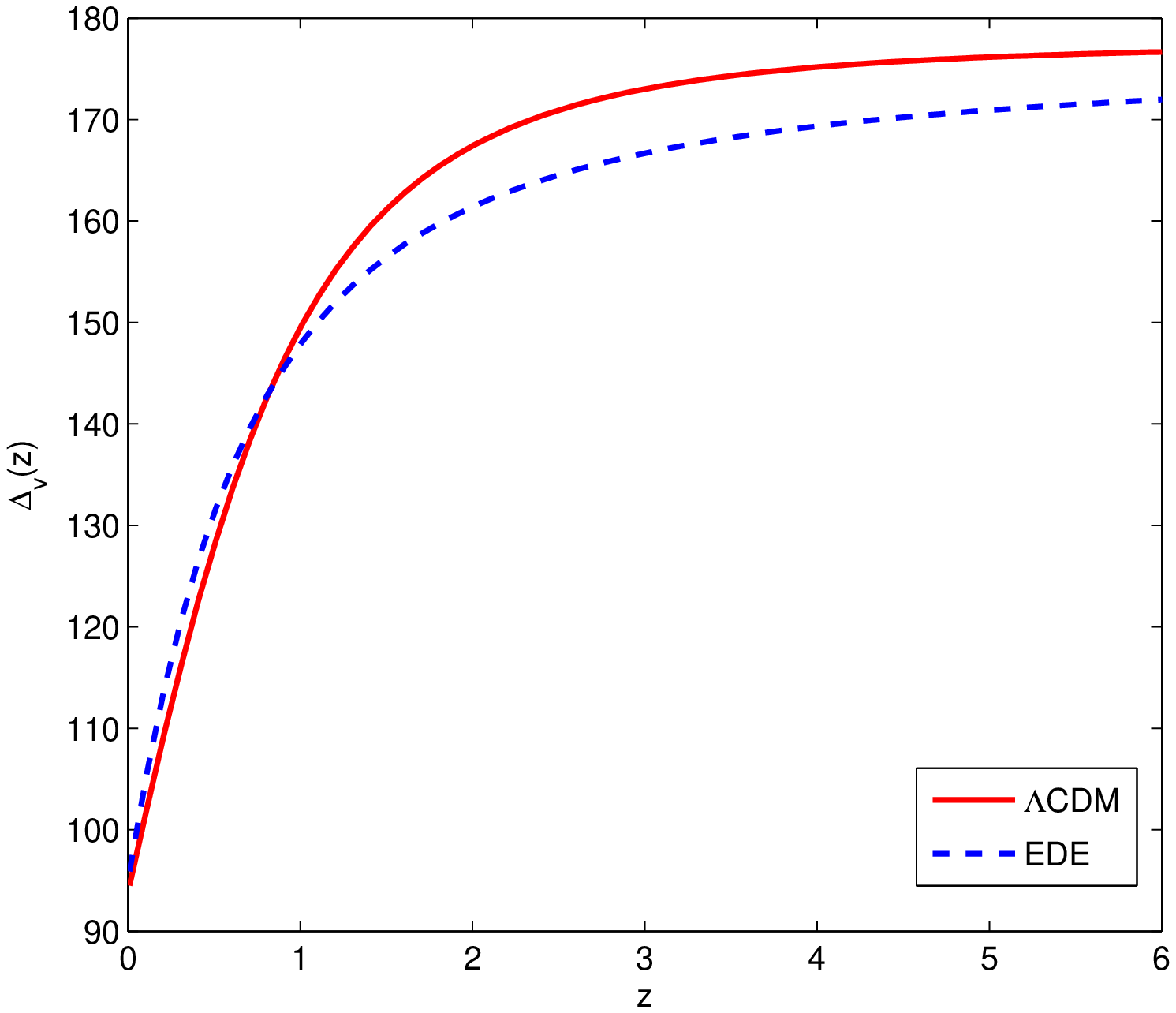, height=7cm, width=8.cm, clip=}
\caption{Numerical results for the linear growth factor, critical density for spherical collapse, and the overdensity at virialisation are illustrated in the upper left-hand, upper right-hand, and bottom panels, respectively, for models I (continuous) and II (dashed).}
\label{fig:evo}
\end{figure}

Numerical results for the linear growth factor, critical overdensity for collapse, and overdensity at virialisation in the EDE model are compared with the corresponding results for the standard 
$\Lambda$CDM model in Fig.~\ref{fig:evo}.

\section{ACKNOWLEDGMENT}

We thank the referee for useful comments. Work at Tel Aviv University 
is supported by a grant from the Israel Science Foundation. 

\section{REFERENCES}
\def\ref{\par\noindent\hangindent 20pt}

\ref Amara A., Refregier A., 2004, \mn, 351, 375
\ref Avila-Reese V., Colin P., Piccinelli G., Firmani C., 2003, \apj, 
598, 36
\ref Bardeen J.M., Bond J.R., Kaiser N., \& Szalay A.S., 1986,\apj, 304,15
\ref Bartelmann M., Doran M., Wetterich C., 2006, \aa, 454, 27
\ref Basilakos S., Voglis N., 2007, \mn, 374, 269
\ref Caldwell R.R., Doran M., M\"{u}ller C.M., Sch\"{a}fer G., Wetterich C., 2003, \apj, 591, L75
\ref Coles P., Barrow J.D., 1987, \mn, 228, 407
\ref Cooray A., 2000, \prd, 62, 103506
\ref Dawson K.S., Holzapfel W.L., Carlstrom J.E., Joy M., LaRoque S.J., Miller A.D., Nagai D., 2002, \apj, 581, 86
\ref Doran M., Lilley M., Schwindt J., Wetterich C., 2001a, \apj, 559, 501
\ref Doran M., Schwindt J., Wetterich C., 2001b, \prd, 64, 123520 
\ref Evrard A.E. et al., 2007, preprint (astro-ph/0702241)
\ref Ferreira P.G., Joyce M., 1998, \prd, 58, 023503
\ref Horellou C., Berge J., 2005, \mn, 360, 1393
\ref Komatsu E., Kitayama T., 1999, \apj, L1
\ref Komatsu E., Seljak U., 2002, \mn, 336, 1256
\ref Komatsu E. et al., 2003, \apjs, 148, 119
\ref Koyama K., Soda J., Taruya A., 1999, \mn, 310, 1111 
\ref Kuo C.L. et al., 2004, \apj, 600, 32
\ref Magliocchetti M., Silva L., Lapi A., De Zotti G., Granato G.L., 
Fadda D., 2006, preprint (astro-ph/0611409)
\ref Majumdar S., 2001, \apj, L7
\ref Maor I., Lahav O., 2005, \jca, 7, 3
%\ref Mason B., Pearson T.J., Readhead A.C.S., Sheperd M.C., Sievers J.L.,
%Udomprasert P.S., Cartwright J.K., Farmer A.J., et al., 2003, \apj, 591, 540
\ref Matarrese S., Verde L., Jimenez R., 2000, \apj, 541, 10
\ref Mathis H., Diego J.M., Silk J., 2004, \mn, 353, 681
\ref Mo H.J., White S.D.M., 1996, \mn, 282, 347
\ref Peacock J.A., Dodds S.J., 1996, \mn, 280, 19
\ref Peebles P.J.E., 1997, \apj, 483, L1
\ref Peebles P.J.E., 1999a, \apj, 510, 523
\ref Peebles P.J.E., 1999b, \apj, 510, 531
\ref Press W.H., Schechter P., 1974, \apj, 187, 425
\ref Readhead A.C.S. et al., 2004, \apj, 609, 498
\ref Refregier A., Teyssier R., 2002, \prd, 66, 043002
\ref Robinson J., Baker J.E., 2000, \mn, 311, 781
\ref Robinson J., Gawiser E., Silk J., 2000, \apj, 532, 1
\ref Roychowdhury S., Ruszkowski M., Nath B.B., 2005, \apj, 634
\ref Sadeh S., Rephaeli Y., Silk J., 2006, \mn, 368, 1583
\ref Spergel D.N., Bean R., Dor\'{e} O., et al., 2006, 
preprint (astro-ph/0603449)
\ref Venemans B.P., Rottgering H.J.A., Miley G.K., van Breugel W.J.M., De Breuck C., Kurk J.D., Pentericci K., Stanford S.A., Overzier R.A., Croft S., Ford H., 2006, preprint (astro-ph/0610567)
\ref Wang P., 2006, \apj, 640, 18
\ref Wetterich C., 2004, \plb, 594, 17
\ref Zeng D-f, Gao Y-h, 2005a, preprint (astro-ph/0412628)
\ref Zeng D-f, Gao Y-h, 2005b, preprint (astro-ph/0505164)

\bsp
\label{lastpage}

\end{document}